# Experimenting Touchless Gestural Interaction for a University Public Web-based Display

Giorgia Persichella*
Master degree in Computer Science
University of Turin, Italy
giorgia.persichella@edu.unito.it

Calogero Luca Lomanto*
Master degree in Computer Science
University of Turin, Italy
cologero.lomanto@edu.unito.it

Claudio Mattutino
Computer Science Department
University of Turin, Italy
claudio.mattutino@unito.it

Fabiana Vernero
Computer Science Department
University of Turin, Italy
fabiana.vernero@unito.it

Cristina Gena
Computer Science Department
University of Turin, Italy
cristina.gena@unito.it

**ABSTRACT**

Interest in and development of touchless gestural interfaces has recently exploded, fueled by the diffusion of both commercial mid-air gesture platforms and public interactive displays. This paper focuses on an application based on Microsoft Kinect that allows students to browse a university website, hosted on a public display, through simple gestures. We present two empirical evaluations where we evaluated how users react to this new way of interaction. In addition to confirming the current lack of standards, our results provide some inspiration for the design of touchless interaction.

**ACM Reference Format:**
Giorgia Persichella*, Calogero Luca Lomanto*, Claudio Mattutino, Fabiana Vernero, and Cristina Gena . 2019. Experimenting Touchless Gestural Interaction for a University Public Web-based Display. In *Proceedings of (CHITALY 2019).* ACM, New York, NY, USA, 5 pages. https://doi.org/10.475/123_4

## 1 INTRODUCTION

In recent years, interest in touchless gestural interfaces has exploded, fueled by the diffusion of commercial mid-air gesture platforms such as Microsoft Kinect, Leap Motion and Myo Armband, of commercial robots, toys and other devices controllable through simple (hand) gestures, and also of other applications as hand-gesture controlled computer screens and on board screens in vehicles, etc. In the meantime, large interactive displays [1] appear in public locations, such as museums, shopping centers, universities, etc.,

* Student who developed the project.

Interactive touchless gestures may be defined as body movements that can be recognized through motion sensing input devices [3]. While gestural and touchless interaction is a natural way of communication that can diminish the barrier between users and interfaces [5], it is still important to contribute with core design principles to enable an intuitive and natural user experience. In fact, standard and shared practices to guide the design of interfaces based on mid-air gestures are still missing [2].

The here-described project was born with the intent of studying and experimenting gestural touchless interaction between users and a public situated display that will be hosted at our University Campus, and will not be directly reachable through touch-based or voice-based interaction. More specifically, we will offer students a Kinect Browser-based version of one of the University web sites presenting a degree course in ICT.

To study user behavior and, in particular, to assess the intuitiveness of gestural input, we carried out a set of empirical evaluations: the basic idea is that users who want to carry out a certain task (e.g., zooming a page) should be able to perform the correct gesture even with no specific training or competencies.

In the following, we will first discuss related work (Section 2). Then, we will present the first evaluation and its results in Section 3, as well as the second evaluation in Section 4. We will conclude the paper in Section 5.

## 2 RELATED WORK

The definition of a vocabulary of intuitive and meaningful gestures is a key point in the design of gestural interfaces. Gentile et al. [4] study such vocabularies comparing the *de facto* standard, the Microsoft Human Interface Guidelines (HIG), with a custom designed interface based on an avatar continuously replaying users' gestures and evaluate how the two interfaces address the interaction blindness issue. Grandhi et al. [6] aim at providing guidelines for the design and implementation of gesture vocabularies for touchless interaction: they present users with pairs of pictures showing a before and after scenario related to simple computer tasks, camouflaged as everyday non-computer scenarios to minimize the influence of pre-existing conceptual models.

A peculiar approach to the definition of intuitive input are guessability studies, where users themselves are asked to suggest appropriate vocabularies. Wobbrock et al. [11] defined guessability as *that quality of symbols which allows a user to access intended referents via those symbols despite a lack of knowledge of those symbols*. In the context of HCI, symbols can be understood as the gestures, commands, buttons or menu items that are used to control an artifact, while referents correspond to the functions and features that can be operated through such symbols.

Guessability studies such as [10] and [8] bear some similarity to our evaluations. Both studies describe a multimodal interaction (speech and gestures) and exploit an elicitation study for web browsing on a living room TV, testing an higher number of functionalities (namely, 15) than we did. In particular Nebeling et al. [10] also developed a Kinect Browser for their study. What is interesting, however, is that their final conclusions are quite similar to ours: in fact, they also found that users assumed their hands would be tracked as when using well-known devices such as mouses. Consequently, suggested interactions were based on such paradigm.

## 3 THE FIRST EMPIRICAL EVALUATION

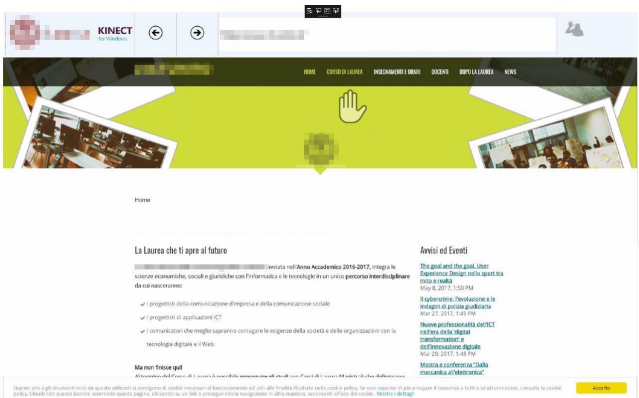

**Figure 1: The Kinect Browser interface.**

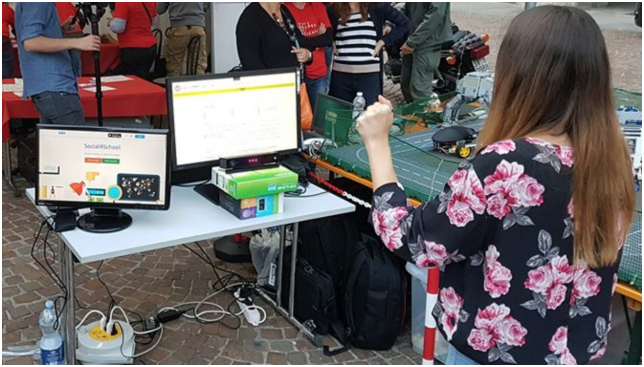

**Figure 2: A user interacting during the first empirical evaluation**

The application under study was realized as a Kinect Browser[1], i.e., a multimodal web browser supporting common browser function following the Microsoft HIG (Human Interface Guidelines) but trying somehow to accommodate the user's preferential choices . The interface of the Kinect Browser is shown in Fig. 1 with the home page of the web site under evaluation. The application is presented in a single main page consisting of a fixed header and a dynamic body. The header of the Kinect Browser contains the back and forth buttons, the navigation bar displaying the URL, and finally a small box showing the user in front of the Kinect sensor, the so-called *User Viewer control*. The body instead includes the real browser wherein the user can browse web pages that change dynamically based on user interaction. Our implementation sup- ported the following basic browser functionalities: "Click Link", press the "Back" and "Forth" Button, "Zoom In and Out" and "Scroll", all realized according to the Microsoft HIG, apart for the "Click Link" functionality, which has not been tested in this study. The home page of the web site under evaluation is shown in Fig. 1.

### 3.1 Method and results

In this first study, our aim was to test the intuitiveness of some Microsoft HIG based gestures, implemented in our prototype.

**Apparatus and Materials**. We carried out our empirical evaluation during the European Researchers' Night[2], which took place in a public square of our city. Our Department had a demonstration stand that hosted also our platform, consisting of a 32-inch display, the Microsoft Kinect sensor, and a desktop computer on which our application was running.

**Participants**. The participants who voluntarily took part in the trial were 14, 7 females and 7 males, all right handed, all without physical constraints, aged 11-46 (mean: 21), with a majority of high school students aged between 10 and 15 (6 subjects, 43%), followed by university students or newly graduated (5 subjects, 36%), and 21% (3 subjects) of graduated workers. Only one participant had had previous experience with the Kinect.

**Procedure**. The tasks proposed to the participants were very general ones, such as: (1) *Finding the correct initial approach to start the interaction*; Interacting with the display and the content of the web page by performing: (2) *Scrolling*, (3) *Zooming*, and (4) *Pressing Back and Forth Buttons*. Participants were given written instructions explaining the goal of the study (i.e., to help us identify natural forms of touchless gestural interaction for browsing the web), then they had to fill in a pre-test asking for their socio-demographic data and their previous experience with mid-air gesture interaction, and finally they could read the four experimental tasks described above. For all the tasks, participants were asked to perform the gesture they would use to carry out the required task on the proposed prototype, in conditions of total autonomy and without any instructions or help. In fact, the goal of the study was to observe participants' spontaneous behaviour. Participants were allowed to try different gestures, until they either guessed the right one, or asked us for help. At the end of the test we carried out an individual

---

[1] The Kinect browser has been realized using the Microsoft environment: the Kinect SDK and the Toolkit. In particular it is a WPF Application developed on Visual Studio making use of C# for the functional part and XAML for the graphic part.
[2] It is an initiative promoted by the European Commission since 2005 which involves thousands of researchers and research institutions each year in all European countries (http://nottedeiricercatori.it).

semi-structured interview to collect user's feedback.

**Results and Discussion**. In our interpretation of the results, we considered as "correct" the gestures that exactly simulated the implemented HIG-based gestures, thus activating the expected behaviour. Regarding task (1), the implemented HIG-based gesture consists in raising one's arm and bringing it over one's head. In this way the Kinect Cursor is enabled and appears on the screen. In general, participants have experienced difficulties in identifying this gesture (excluding the experienced one, 8 out of 13 participants (69.2%) failed this task[3]). In fact, participants pointed out that, although this gesture is not difficult to perform, it does not appear natural and does not conform to their habits. In contrast, most participants have tried to "be seen" by the sensor by waving hello and/or approaching the sensor with their whole body or placing their hands near the Kinect camera. These behaviors can be partly explained by the daily interaction they have with their touch devices, and their instinctive need for some contact with interactive media, and partly by the implicit social norms evoked by this task related to the beginning of an interaction/conversation.

To complete task (2), the scrolling task, participants first had to better understand the potentiality of the Kinect Cursor. A first exploration phase was therefore necessary, where participants experimented with different hand movements in the body part of the application. Hence, participants were able to learn one of the fundamental states of the cursor: the gripping. Most participants have been able to identify the gripping gesture after a limited time, probably thanks to the feedback provided by the browser, which shows the closed fist to suggest the "gripped" condition. Once the gripping gesture was learnt, most participants exploited it to complete the scrolling task (12 out of 14, equal to 85.7%).

Notice that the order of the assigned tasks somehow influenced the rest of the interaction. After task (2), participants were able to grip the page, an essential ability to carry out zooming, as required in task (3). According to HIG, in fact, zooming is done by gripping the page and moving one's fist back and forth along the Z-axis. However, the zooming task was not intuitive: in fact, it cannot be mapped to similar ones learnt in touch screens, where "pinch" gestures would be used instead. Therefore, suggestions were made after a while to allow participants to successfully complete the task. Even with the suggestions, 4 participants failed the task (28.6%).

The last task assigned to participants was to press the back and forth buttons (see Fig. 1) (task (4)). Most participants tried to exploit similarities with the gestures they had learnt up to that moment. It is interesting to note that almost all of them (11 out of 14, 78%) first suggested to "press" the buttons through the "gripping gesture", closing their hands in a fist near the button in the interface. This last aspect makes it clear how participants continually refer to their previous experience to perform unknown tasks. Some participants (7 out of 14, 50%) also tried to map gestural interaction with the interaction with a physical mouse, where they would press the left mouse button. Finally, the standard, HIG-based gesture associated with this task (i.e., opening one's hand and simulating a pressure towards the screen (*push-to-press* activation gesture) was only suggested by 6 out of 14 participants (43%), the ones in the age range 10-15, thus confirming the difficulty of this gesture, as already observed in [4].

To sum up, we noticed that participants in general have shown problems with task (1), (3) and (4), but participants with an age range between 10 and 15 have more quickly adapted their behavior to the application requirements, managing to interact and obtaining better results. Users with these features are probably more open to new experiences, as they are able to learn much faster, providing an excellent prospect of expanding technological innovation in the future. The past experience of participants acquired during task execution, as well as their experience in desktop and mobile interaction, have certainly influenced their performance.

Furthermore, participants spontaneously performed HIG-based gestures only occasionally, thus implying that they are not very intuitive. When highly intuitive gestures cannot be used (for example because of technical constraints), users should be provided with specific hints. In [8], the authors propose to give precise indications about the interactions, by means of video tutorials or posters, while in [4] they propose to use an avatar suggesting the right gesture to be performed in front of the display, also to overcome the problem of the interaction blindness, namely the inability of the public to recognize the interactive capabilities of a certain surface.

## 4 THE SECOND EMPIRICAL EVALUATION

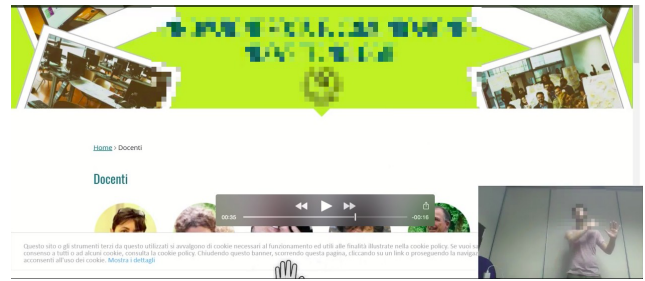

**Figure 3: A user interacting during the second empirical evaluation**

In the second version of the application, we decided to extend our set of gestures beyond the possibilities offered by Microsoft HIG and to consider also other tasks, such as those described in [7]. We decided to keep the interface of the Kinect Browser (see Fig. 3), but making some changes. In this version of the prototype, the header contains only the navigation bar for displaying the URL and a small box representing the segment of the body detected by Kinect. Since the new implementation aims at replacing the "back" and "forth" buttons with two gestures, we removed the corresponding buttons. As for the body part of the application, no changes were made.

In order to add new gestures, we exploited the Visual Gesture Builder tool offered by the Kinect Studio Tool. It exploits machine learning algorithms in order to learn the new gestures, taking advantage of a support database where new clips, created with Kinect Studio, are stored. The clips contain data about the newly introduced gestures, and these will be recognized once the tagging process has taken place. Unlike the previous application, both hands can be used: the right hand for performing the gestures, and the left hand for controlling the cursor. We have identified 7 actions taking inspiration from [7]:

> **Start**: This is a dynamic gesture used to start the communication with the web browser. The gesture mimics a simple greeting;
> **Click**: This gesture mimics a click on a link with the closed

---

[3] Notice that tasks were considered successful when the participants identified the right gesture and could therefore complete the required task.

(left) hand performing the gripping gesture near the link;
- **Back**: This gesture involves a movement of the right hand along the x-axis, from right to left at chest level, in order to go back in the browser history;
- **Forth**: This gesture, similar to the previous one, involves moving the right hand along the x-axis but, this time, from left to right, in order to go forward in the browser history;
- **Minimize**: This gesture is used to minimize the page, and involves moving the right hand along the y-axis from top to bottom;
- **Maximize**: This gesture is used to re-open a minimized page, and involves moving the right hand along the y-axis from bottom to top;
- **Reload**: This is a dynamic gesture used to reload the page, and involves moving the hand clockwise.

In addition, we also tested the **zoom** action, maintaining the same gesture proposed in the previous experiment.

## 4.1 Method and results

The goal of this second study was to test the intuitiveness of the above implemented gestures. This time we also considered, as metric, the successful attempts after the first (wrong) ones.

**Apparatus and Materials**. The evaluation was carried out in our HCI lab. Participants were standing in front of a 32-inch display connected to a Microsoft Kinect sensor and to the desktop computer where our application was running.

**Participants**. We involved 14 people. All of them were computer science students, 12 males and 2 females, aged between 20 and 30, with an average age of 22.9, SD=1.94. Only 4 participants (28%) had already had some experience with the Microsoft Kinect, or with other motion sensing devices.

**Procedure**. No specific information on the correct gestures to perform was provided. However, we explained to each participant that, in order to perform the gestures, both hands should be used, with the left hand being used to control the mouse, and the right hand being used to operate the main browser functionalities. As in our previous evaluation, the tasks proposed to the participants were very general ones, such as: Using a gesture (1) start the communication with the web application; (2) click on any link on the page; (3) access the previous page in the browser history, as you do when you click on the back button; (4) access the following page in the browser history, as you do when you click on the forth button; (5) try to minimize the page; (6) try to maximize the page; (7) try to zoom-in/zoom-out the page; (8) try to reload the page. Participants were allowed to try different gestures, until they either guessed the right one, or asked us for help.

**Results and Discussion**. Similar to what we did in our previous evaluation, we considered as "correct" the gestures that exactly simulated the required gestures, thus activating the expected behaviour.

Regarding task (1), only 6 participants (42.86%) performed the correct start gesture, while most of them (8 participants, 57.14%) did not. Considering only the participants who were unable to perform the correct gesture on their first attempt, however, 50% arrived at the solution after a maximum of 2 attempts. The first gestures for the most part either corresponded to the pressure of a button with one's hand, or imitated moving the finger of the hand forward the mouse click.

Also for task (2), 6 participants (42.86%) were able to correctly click on links, while the remaining 8 (57.14%) were not. Of these last ones, however, 50% of them arrived at the solution after maximum 2 attempts.

Regarding task (3), the back action, 11 people (79%) performed the correct gesture. In contrast, the remaining 3 (21%) tried to make gestures that included only the use of their hand and not of their entire arm, for example tilting their hand from left to right.

Regarding task (4), the forth action, the correct gesture was identified by all the 14 participants (100%). The reason for this success probably lies in the fact that this task immediately followed task (3), where participants experimented with the opposite action. Hence, all the participants logically performed the opposite gesture with respect to the previous one.

Regarding task (5), the minimize action, 11 participants (79%) correctly identified the gesture. The remaining ones mostly tried to minimize the page by closing their fists, and they did not guess the right gesture.

Curiously, regarding task (6), the maximize action, only 12 participants out of 14 (85%) identified the correct gesture.

Regarding task (7), the zoom action, all the 14 participants (100%) failed to identify the correct gesture, even after several attempts. In contrast, they all mimicked a gesture that involved the symmetrical movement of their hands from the center outwards to zoom in, and the opposite gesture to zoom out.

As for task (8), the reload action, 6 participants (42.86%) performed the correct gesture, while the other 8 (57.14%) tried different gestures, among which we were unable to identify any common pattern.

## 5 CONCLUSION

Our empirical evaluations confirmed the lack of standard vocabularies consisting of highly intuitive gestures for touchless interaction. We acknowledge the limitation of our studies due to few people involved, however they offered results close to the ones of other similar studies, above mentioned. Drawing on the discussion of our results, two different approaches can be adopted in order to tackle this open problem. On the one hand, touchless gestural interfaces might include specific features aimed at suggesting the correct gestures, such as avatars, video tutorials or posters. On the other hand, sets of intuitive gestures could be identified through guessability studies [10], and bearing in mind the fact that people usually draw on their previous experience, searching for similarities, when they have to perform new gestures. Thus, for example, we observed that using opposite gestures for opposite actions drastically reduced the number of errors.

# REFERENCES

[1] Carmelo Ardito, Paolo Buono, Maria Francesca Costabile, and Giuseppe Desolda. 2015. Interaction with Large Displays: A Survey. *ACM Comput. Surv.* 47, 3 (2015), 46:1–46:38. https://doi.org/10.1145/2682623

[2] Arthur Theil Cabreira and Faustina Hwang. 2015. An analysis of mid-air gestures used across three platforms. In *Proceedings of the 2015 British HCI Conference, Lincoln, United Kingdom, July 13-17, 2015.* 257–258. https://doi.org/10.1145/2783446.2783599

[3] Alessandro Carcangiu. 2017. A Declarative and Classifier Gesture Recognition Method for Creating an Effective Feedback and Feedforward System. In *Proceedings of the Doctoral Consortium, Posters and Demos at CHItaly 2017 co-located with 12th Biannual Conference of the Italian SIGCHI Chapter (CHItaly 2017), Cagliari, Italy, September 18-20, 2017.* 13–24. http://ceur-ws.org/Vol-1910/paper0102.pdf

[4] Vito Gentile, Salvatore Sorce, Alessio Malizia, Dario Pirrello, and Antonio Gentile. 2016. Touchless Interfaces For Public Displays: Can We Deliver Interface Designers From Introducing Artificial Push Button Gestures?. In *Proceedings of the International Working Conference on Advanced Visual Interfaces, AVI 2016, Bari, Italy, June 7-10, 2016.* 40–43. https://doi.org/10.1145/2909132.2909282

[5] Sukeshini A. Grandhi, Gina Joue, and Irene Mittelberg. 2011. Understanding Naturalness and Intuitiveness in Gesture Production: Insights for Touchless Gestural Interfaces. In *Proceedings of the SIGCHI Conference on Human Factors in Computing Systems (CHI '11)*. ACM, New York, NY, USA, 821–824. https://doi.org/10.1145/1978942.1979061

[6] Sukeshini A. Grandhi, Gina Joue, and Irene Mittelberg. 2011. Understanding Naturalness and Intuitiveness in Gesture Production: Insights for Touchless Gestural Interfaces. In *Proceedings of the SIGCHI Conference on Human Factors in Computing Systems (CHI '11)*. ACM, New York, NY, USA, 821–824. https://doi.org/10.1145/1978942.1979061

[7] Anthony Jameson, Silvia Gabrielli, Per Ola Kristensson, Katharina Reinecke, Federica Cena, Cristina Gena, and Fabiana Vernero. 2011. How can we support users' preferential choice? In CHI '11 Extended Abstracts on Human Factors in Computing Systems (CHI EA '11<). Association for Computing Machinery, New York, NY, USA, 409–418. https://doi.org/10.1145/1979742.1979620

[8] Meredith Ringel Morris. 2012. Web on the Wall: Insights from a Multimodal Interaction Elicitation Study. In *Proceedings of the 2012 ACM International Conference on Interactive Tabletops and Surfaces (ITS '12)*. ACM, New York, NY, USA, 95–104. https://doi.org/10.1145/2396636.2396651

[9] Jörg Müller, Robert Walter, Gilles Bailly, Michael Nischt, and Florian Alt. 2012. Looking Glass: A Field Study on Noticing Interactivity of a Shop Window. In *Proceedings of the SIGCHI Conference on Human Factors in Computing Systems (CHI '12)*. ACM, New York, NY, USA, 297–306. https://doi.org/10.1145/2207676.2207718

[10] Michael Nebeling, Alexander Huber, David Ott, and Moira C. Norrie. 2014. Web on the Wall Reloaded: Implementation, Replication and Refinement of User-Defined Interaction Sets. In *Proceedings of the Ninth ACM International Conference on Interactive Tabletops and Surfaces, ITS 2014, Dresden, Germany, November 16 - 19, 2014.* 15–24. https://doi.org/10.1145/2669485.2669497

[11] Jacob O. Wobbrock, Htet Htet Aung, Brandon Rothrock, and Brad A. Myers. 2005. Maximizing the Guessability of Symbolic Input. In *CHI '05 Extended Abstracts on Human Factors in Computing Systems (CHI EA '05)*. ACM, New York, NY, USA, 1869–1872. https://doi.org/10.1145/1056808.1057043